\documentclass[fleqn,10pt]{SelfArx} 

\usepackage[english]{babel} 

\usepackage{lipsum} 
\usepackage{afterpage}

 \definecolor{color1}{RGB}{0,0,110} 
\definecolor{color2}{RGB}{0,20,80} 

\usepackage{hyperref} 
\hypersetup{hidelinks,colorlinks,breaklinks=true,urlcolor=color2,citecolor=color1,linkcolor=color1,bookmarksopen=false,pdftitle={Title},pdfauthor={Author}}

\JournalInfo{Uploaded to arXiv.org} 
\Archive{} 

\PaperTitle{Controlling ion transport through nanopores: modeling transistor behavior} 

\Authors{Eszter M\'adai\textsuperscript{1}, Bart\l{}omiej Matejczyk\textsuperscript{2}, Andr\'as Dallos\textsuperscript{1}, M\'onika Valisk\'o\textsuperscript{1},  Dezs\H{o} Boda\textsuperscript{1,3}*} 
\affiliation{\textsuperscript{1}\textit{Department of Physical Chemistry, University of Pannonia,  P. O. Box 158, H-8201 Veszpr\'em, Hungary}} 
\affiliation{\textsuperscript{2}\textit{Department of Mathematics, University of Warwick, CV4 7AL Coventry, United Kingdom}} 
\affiliation{\textsuperscript{3}\textit{Institute of Advanced Studies K\"oszeg (iASK), Chernel u. 14, H-9730 K\"oszeg, Hungary}} 
\affiliation{*\textbf{Corresponding author}: dezsoboda@gmail.com} 

\newcommand{\keywordname}{Keywords} 

\Abstract{We present a modeling study of a nanopore-based transistor computed by a mean-field continuum theory (Poisson-Nernst-Planck, PNP) and a hybrid method including particle simulation (Local Equilibrium Monte Carlo, LEMC) that is able to take ionic correlations into account including finite size of ions.
	The model is composed of three regions along the pore axis with the left and right regions determining the ionic species that is the main charge carrier, and the central region tuning the concentration of that species and, thus, the current flowing through the nanopore.
	We consider a model of small dimensions with the pore radius comparable to the Debye-screening length ($R_{\mathrm{pore}}/\lambda_{\mathrm{D}}\approx 1$), which, together with large surface charges provides a mechanism for creating depletion zones and, thus, controlling ionic current through the device.
	We report scaling behavior of the device as a function the $R_{\mathrm{pore}}/\lambda_{\mathrm{D}}$ parameter.
	Qualitative agreement between PNP and LEMC results indicates that mean-field electrostatic effects determine device behavior to the first order. }

\begin{document}

\flushbottom 

\maketitle 


\thispagestyle{empty} 


\section{Introduction}
\label{sec:intro}

In this work, we present a modeling study for a nanopore-based transistor through which ionic current can be controlled via manipulating the charge pattern on the pore wall.
Nanopores are nano-scale holes in synthetic membranes made of, for example, silicon, graphene, or plastic \cite{abgrall_book,iqbal_book_2011}. 
Their natural relatives are called ion channels that, as parts of a complex physiological machinery, facilitate controlled ion transport through the cell membrane \cite{Hille}.
Similarly, synthetic nanopores are essential building blocks of nanofluidic circuitries that aim to control the behavior of fluids in the nanometer scale \cite{abgrall_book}.
Furthermore, nanopores in proton-conducting membranes are the key elements for microfuel cells, which can convert chemical energy directly into electricity \cite{abgrall_book}, most effectively from alternative fuels that has very high energy content by weight because of the high hydrogen content in their molecular structures.

The radius of our nanopore model is comparable to the characteristic screening length of the electrolyte leading to behavior different from micropores such as formation of extended depletion zones and correlations between neighboring regions.
Electrostatic and excluded volume correlations between ions are expected to be important requiring special computational techniques beyond the mean-field level of the Poisson-Nernst-Planck (PNP) theory.
We use a Monte Carlo (MC) particle simulation method in comparison with PNP to compute these correlations and to assess the applicability of PNP in such nano-scale confinements.
We show that transistor behavior scales with the pore radius relative to the screening length, thus making it possible to design nanofluidic devices of varying length scales by adjusting the electrolyte concentration applied in the device.

Nanopores with controlled ion flow are related to semiconductor transistors that modulate the current between the emitter and collector via tuning the availability of charge carriers in a certain region of the device.
This control can be realized by injecting charge carriers or by manipulating the electric field, and, thus, the probability of charge carriers inhabiting the device (e.g., their concentrations).
Control of electric field is commonly done by setting the electrical potential at a third electrode, the gate.

In nanopores, there are several ways to regulate the electric field inside the pore.
Using embedded electrodes, electrically tunable nanopore-based transistors can be fabricated
\cite{burgmayer_jacs_1982,Fan_PRL_2005,karnik_nl_2005,karnik_apl_2006,horiuchi_loac_2006,gracheva_acsnano_2008,kalman_am_2008,Kalman_BJ_2009,cheng_acsnano_2009,tybrandt_natcomm_2012,Lee_NS_2015,Fuest_nl_2015,Fuest_AC_2017} similarly to field-effect semiconductor transistors.
Control of cation/anion concentration in nanopores, however, can also be achieved by manipulating the surface charge pattern on the wall of the nanopore \cite{kuo_l_2001,stein_prl_2004,Siwy_2004,singh_jap_2011,jiang_pre_2011,singh_sab_2016,Singh_PCCP_2017} just as the density of electrons and holes can be controlled with doping in semiconductor devices.
Moreover, if the membrane is made of a semiconductor material, the two techniques can be combined \cite{gracheva_nl_2007,gracheva_acsnano_2008,Gracheva_JCE_2008,melnikov_bd_dna_2012,nikolaev_jce_2014}, namely, ion accumulation can be controlled both with doping the semiconductor material confining the pore and with the electrical potential imposed on it.

The charge of the nanopore can be manipulated with chemical methods by anchoring functional groups to the pore wall \cite{siwy_prl_2002,Kim_2010,siwy_csr_2010,Ali_ACSnano_2012,gibb_chapter_2013,Duan_bmf_2013,guan_nanotech_2014,Fuest_AC_2017}.
Surface charge can also be modulated with pH if protonation/deprotonation of the functional groups is pH sensitive \cite{wang_rm_2009,Xue_2014,Lin_2015}.
Sensors can be designed if molecules attached to the pore wall can bind other molecules selectively so that binding these molecules modulates the current of the background elecrolyte through the pore in a characteristic and identifiable way \cite{sexton_mbs_2007,gyurcsanyi_trac_2008,howorka_csr_2009,vlassiouk_jacs_2009,piruska_csr_2010,fahie_acs_2016,Ali_JACS_2011,Ali_Lang_2017,Ali_AC_2018,perezmitta_small_2018}.  
In the special case of charged biopolymers (such as DNA), electric current and/or field changes during the translocation of the polymer through the nanopore that can make sequencing possible \cite{Ai_AC_2010,otto_chapter_2013}.

The difference between nanopores and their microfluidic counterparts \cite{vandenberg_csr_2010} is that the radial dimension of the nanopore (pore radius, $R_{\mathrm{pore}}$) is comparable to the Debye screening length, $\lambda_{\mathrm{D}}$ of the electrolyte that is defined as
\begin{equation}
\lambda_{\mathrm{D}} = \left( \sum_{i} \dfrac{q_{i}^{2}c_{i}}{\epsilon_{0}\epsilon kT} \right)^{-1/2},
\label{eq:lambdaD}
\end{equation} 
where $q_{i}$ is the charge and $c_{i}$ is the bulk concentration of ionic species $i$, respectively, $\epsilon_{0}$ is the permittivity of vacuum, $\epsilon$ is the dielectric constant of the electrolyte, $k$ is Boltzmann's constant, and $T$ is the absolute temperature.
The Debye-length is loosely identified with the thickness of the electrical  double layer that is formed by the ions near the charged wall of the nanopore.
If the pore is narrow enough or the double layer is wide enough (at lower concentration), the diffuse part of the double layer extends into the pore preventing the formation of a bulk electrolyte in the central region of the pore \cite{Abgrall_2008,daiguji_csr_2010,bocquet_csr_2010}.

The double layers are generally depleted of the coions whose charge has the same sign as the pore's surface charge.
If we apply large enough surface charge, depletion zones (regions of  very low local concentration) for the coions can be established in certain sections of the nanopore. 
Since consecutive sections of the nanopore along its axis behave as resistors connected in series, any of these elements with a deep depletion zone (with large resistance) makes the resistance of the whole pore large.
Transistor behavior can be produced if we can adjust the depth of the depletion zone in such a section.

This paper explores transistor behavior for a model nanopore for different charge patterns, pore geometry, and bulk concentration.
We focus on small $R_{\mathrm{pore}}/\lambda_{\mathrm{D}}$ ratios by using relatively small pore radii below 2.5 nm.
In such a narrow pore, we expect that ion size effects are important, therefore, we model ions	as charged hard spheres immersed in a continuum background dielectric.
To compute ionic correlations beyond mean-field, we use the Local Equilibrium Monte Carlo (LEMC) simulation method that is an adaptation of the Grand Canonical Monte Carlo (GCMC) technique to a non-equilibrium situation \cite{boda-jctc-8-824-2012,boda-arcc-2014,boda-jml-189-100-2014,hato-cmp-19-13802-2016,hato-pccp-19-17816-2017,matejczyk-jcp-146-124125-2017,madai-jcp-147-244702-2017}.
We couple this method to the Nernst-Planck (NP) equation (NP+LEMC method) to compute flux just as PNP does.
PNP is a continuum theory that, sometimes combined with the Navier-Stokes equation to describe water flux, is commonly used to study nanopores \cite{daiguji_nl_2004,daiguji_nl_2005,constantin_pre_2007,vlassiouk_nl_2007,kalman_am_2008,vlassiouk_acsnanno_2008,cheng_acsnano_2009,pietschmann_pccp_2013,singh_jap_2011b,singh_lc_2012,singh_pccp_2016,singh_sab_2016,Singh_PCCP_2017,cervera_epl_2005,cervera_jcp_2006,ramirez_jcp_2007,cervera_ea_2011,gracheva_acsnano_2008,Gracheva_JCE_2008,nikolaev_jce_2014,pardon_acis_2013,volkov_l_2014,park_mfnf_2015,tajparast_bba_2015}.
PNP studies that consider transistor models similar to ours will be discussed in the Discussions in relation to our results \cite{gracheva_acsnano_2008,Gracheva_JCE_2008,nikolaev_jce_2014,park_mfnf_2015,daiguji_nl_2005,singh_jap_2011b,singh_lc_2012,Singh_PCCP_2017}.

This work belongs to a series of publications \cite{hato-cmp-19-13802-2016,hato-pccp-19-17816-2017,matejczyk-jcp-146-124125-2017} in which we use a multiscale modeling approach to study nanodevices using models of different resolutions computed by the appropriate computational method. 
In a previous publication \cite{hato-pccp-19-17816-2017}, we compared results for an implicit water (studied by NP+LEMC) and  an explicit water (studied by molecular dynamics) for a bipolar nanopore model (``$-+$'' charge distribution on the pore wall along the axis) and justified the applicability of the implicit water model.
We showed that the reduced model properly captures device behavior (rectification), because it includes those degrees of freedom (ions' interaction with pore, pore charges, and applied field) that are necessary to reproduce the axial behavior of concentration profiles and ignores those (explicit water molecules) that determine the radial behavior.
As it turned out, radial behavior had secondary importance in reproducing the device behavior.

In another study \cite{matejczyk-jcp-146-124125-2017} we showed that even mean-field electrostatics captures those effects that provide the qualitative axial behavior of the bipolar nanopore by comparing PNP to NP+LEMC.
That work justified using PNP in computational studies of nanopores by calibrating PNP to a particle simulation method (LEMC), at least, for the case of 1:1 electrolytes.
In this work, we continue this study by creating a transistor model that can be viewed as two bipolar diodes combined head-to-head (``$-+-$'').
For that reason, it is often called a bipolar transistor.
In this symmetric three-region model, the two ``$-$'' regions are used to define the main charge carrier ionic species (cations), while the central region is used to control the concentration of cations by tuning the surface charge of that region.

\section{Model and methods}
\label{sec:models}

\subsection*{Model of nanopore}
\label{subsec:model}

The device studied here is composed of two baths separated by a membrane.
The two sides of the membrane is connected by a single cylindrical pore that penetrates the membrane. 
The system has a rotational symmetry around the axis of the pore, therefore, the solution is presented in terms of cylindrical coordinates $z$ and $r$ (the simulation cell in the LEMC simulation is three-dimensional, however).
The solution domain is a cylinder of 30 nm width and 9 nm radius for a pore with $H_{\mathrm{pore}}=10$ nm length and $R_{\mathrm{pore}}=1$ nm radius.
For longer and wider pores, these dimensions are proportionately larger.
Fixed values of the concentrations and potential   are prescribed on the half-cylinders on the left and right hand side.

\begin{figure}[t]
	\begin{center}
		\scalebox{0.4}{\includegraphics*{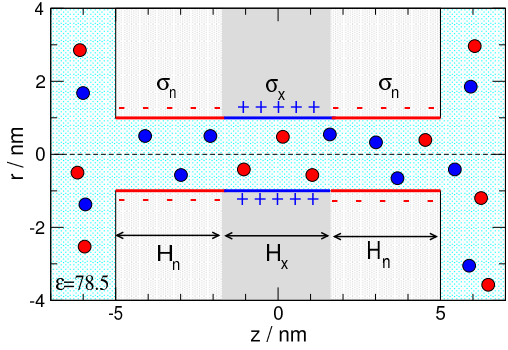}}
	\end{center}
	\caption{Schematics of the cylindrical nanopore that has three regions of lengths $H_{\mathrm{n}}$, $H_{\mathrm{x}}$, and $H_{\mathrm{n}}$. These regions carry $\sigma_{\mathrm{n}}$, $\sigma_{\mathrm{x}}$, and $\sigma_{\mathrm{n}}$ surface charges, respectively. The radius of the nanopore is $R_{\mathrm{pore}}$. The simulation cell is larger than this domain of this figure, but also rotationally symmetric; the three-dimensional model is obtained by rotating the figure about the $z$-axis. The electrolyte inside the pore and on the two sides of the membrane is represented as charged hard sphere ions immersed in a dielectric continuum of dielectric constant $\epsilon=78.5$. The dielectric constant is the same everywhere including the interior of the membrane. The PNP model closely mimics this model as described in the main text.}
	\label{Fig1}
\end{figure}

The membrane and the pore is confined by hard walls.
The thickness of the membrane is the length of the pore, $H_{\mathrm{pore}}$.
A symmetric charge pattern is created on the wall of the nanopore as shown in Fig.\ \ref{Fig1}.
There are regions of widths $H_{\mathrm{n}}$ on the two sides of the pore carrying $\sigma_{\mathrm{n}}$ surface charges, while there is a central region of width $H_{\mathrm{x}}$ and charge $\sigma_{\mathrm{x}}$.

Here, the $\sigma_{\mathrm{n}}$ regions set the main charge carrier.
In this study, we typically use negatively charged regions (hence the notation n), so the main charge carriers are the cations because the $\sigma_{\mathrm{n}}$ surface charges produce depletion zones of anions in these regions.

The task of the central region with the adjustable surface charge $\sigma_{\mathrm{x}}$ is to regulate the flow of cations (this is the independent variable of the device, hence the notation x).
If $\sigma_{\mathrm{x}}$ is positive, it produces a depletion zone for cations, so the pore contains depletion zones for both ionic species.
The total current, therefore, is small.
This corresponds to the OFF state of the device.
We distinguish special cases for combinations of $\sigma_{\mathrm{n}}$ and $\sigma_{\mathrm{x}}$ when these surface charges are -1, 0, or 1 $e/\mathrm{nm}^{2}$.
We denote these charges by symbols ``$-$'', ``$0$'', and ``$+$'', respectively.
So if $\sigma_{\mathrm{n}}=-1$ $e/\mathrm{nm}^{2}$ and $\sigma_{\mathrm{x}}=1$ $e/\mathrm{nm}^{2}$, the nanopore is characterized by the string ``$-+-$'' (as in Fig.\ \ref{Fig1}).

\subsection*{Reduced model of electrolyte and ion transport}

We describe the interactions and transport of ions with a reduced model, in which the degrees of freedom of solvent molecules is coarse-grained into response functions.
Particularly, water is modeled as an implicit continuum background that has two kinds of effect on ions.

Water molecules screen the charges of ions. 
This ``energetic'' effect is taken into account by a dielectric constant, $\epsilon$, in the denominator of the Coulomb potential acting between the charged hard spheres with which we model the ions:
\begin{equation}
u_{ij}(r)
=\left\{
\begin{array}{ll}
\infty & \; \mbox{for} \; \;  r<R_{i}+R_{j}\\
\dfrac{q_{i}q_{j}}{4\pi \epsilon_{0}\epsilon r} & \; \mbox{for} \; \; r \geq R_{i}+R_{j}  ,
\end{array}
\right. 
\label{eq:pm}
\end{equation} 
where $R_{i}$ is the radius of ionic species $i$ and $r$ is the distance between the ions.

Water molecules hinder the diffusion of ions with friction.
This ``dynamic'' effect is taken into account by a diffusion coefficient, $D_{i}(\mathbf{r})$, in the Nernst-Planck (NP) transport equation for the ionic flux:
\begin{equation}
-kT\mathbf{j}_{i}(\mathbf{r}) = D_{i}(\mathbf{r})c_{i}(\mathbf{r})\nabla \mu_{i}(\mathbf{r}),
\label{eq:np}
\end{equation} 
where $\mathbf{j}_{i}(\mathbf{r})$ is the particle flux density of ionic species $i$, $c_{i}(\mathbf{r})$ is the concentration, and $\mu_{i}(\mathbf{r})$ is the electrochemical potential profile.

The diffusion coefficient profile, $D_{i}(\mathbf{r})$, is a parameter to be specified by the user.
In the baths, we can use experimental values.
We can adjust its value inside the pore to experiments (as in the case of the Ryanodine Receptor calcium channel \cite{boda-arcc-2014,fertig-hjic-45-73-2017}) or to results of molecular dynamics simulations (as in the case of bipolar nanopores \cite{hato-pccp-19-17816-2017}).
It can also be just an arbitrary parameter as in this study, because its value inside the pore tunes the current practically linearly and its precise value is inconsequential from the point of view of understanding the behavior of the system. 

To solve the NP equation, we need a closure between the concentration profile, $c_{i}(\mathbf{r})$, and the electrochemical potential profile, $\mu_{i}(\mathbf{r})$.
Such a closure is provided  by statistical mechanics. 
We apply two kinds of methods in this work, a particle simulation method (LEMC) and a continuum theory method (PNP).
Once the relation between $c_{i}(\mathbf{r})$ and  $\mu_{i}(\mathbf{r})$ is available, a self-consistent solution is obtained iteratively in which the conservation of mass, namely, the continuity equation, $\nabla \cdot \mathbf{j}_{i}(\mathbf{r}) = 0$,is satisfied.

\subsection*{Local Equilibrium Monte Carlo}
\label{subseq:lemc}

LEMC is an adaptation of the GCMC technique to a non-equilibrium situation \cite{boda-jctc-8-824-2012,boda-jml-189-100-2014,boda-arcc-2014,fertig-hjic-45-73-2017}.
The independent state function of the LEMC simulation is the chemical potential profile, $\mu_{i}(\mathbf{r})$, while the output variable is the concentration profile, $c_{i}(\mathbf{r})$.
Chemical potential is constant in space in equilibrium for which GCMC simulations were originally designed.
Out of equilibrium, however, $\mu_{i}(\mathbf{r})$ is a space-dependent quantity.

The transition from global equilibrium to non-equilibrium is possible by assuming local equilibrium (LE).
We divide the solution domain into small volume elements, $\mathcal{B}^{\alpha}$, and assume that the chemical potential is constant in this volume, $\mu_{i}^{\alpha}$.
This value tunes the probability that ions of species $i$ occupy this volume.
LEMC applies ion insertion/deletion steps that are very similar to those used in global-equilibrium GCMC with the differences that (1) the electrochemical potential value, $\mu_{i}^{\alpha}$, assigned to the volume element in which the insertion/deletion happens, $\mathcal{B}^{\alpha},$ is used in the acceptance probability, (2) the volume of the volume element, $V^{\alpha}$, is used in the acceptance probability, and (3) the energy change associated with the insertion/deletion, $\Delta U$, contains all the interactions from the whole simulation cell, not only from volume element $\mathcal{B}^{\alpha}$.
A self-consistent solution is found by an iterative process, in which $\mu_{i}^{\alpha}$ is changed until conservation of mass ($\nabla \cdot \mathbf{j}_{i}=0$) is satisfied.
Details can be found in our earlier publications \cite{boda-jctc-8-824-2012,boda-jml-189-100-2014,boda-arcc-2014,fertig-hjic-45-73-2017}. 

The advantage of this technique (coined as NP+LEMC) is that it correctly computes volume exclusion and electrostatic correlations between ions, so it is beyond the mean-field level of the PNP theory. 
Its advantage compared to the Brownian Dynamics method \cite{chung-bj-77-2517-1999,im_bj_2000,berti-jctc-10-2911-2014} is that sampling of ions passing the pore is not necessary: current is computed with the NP equation.
Sampling of passing ions can be poor especially when these events are rare due to the small current associated with the depletion zones of ions.
The transistors studied here belong to this category because their behavior is governed by these depletion zones. 
The NP+LEMC method has been successfully applied for membranes \cite{boda-jctc-8-824-2012,hato-jcp-137-054109-2012}, ion channels \cite{boda-jml-189-100-2014,boda-arcc-2014,hato-cmp-19-13802-2016,fertig-hjic-45-73-2017} and nanopores \cite{hato-pccp-19-17816-2017,matejczyk-jcp-146-124125-2017,madai-jcp-147-244702-2017}.

In the three-dimensional LEMC model, the pore charges are placed on the pore wall as point charges on a grid.
The size of a grid surface element is about $0.2\times 0.2$ nm$^{2}$.
The magnitude of point charges was calculated so that the surface charge density agrees with the preset values $\sigma_{\mathrm{n}}$ or $\sigma_{\mathrm{x}}$.
This solution was chosen to mimic the continuous charge distribution used in the PNP calculations.

\subsection*{Poisson-Nernst-Planck theory}
\label{subseq:pnp}

In this work, we apply a two-dimensional version of the  steady state PNP  system as described in  Matejczyk et al.\ \cite{matejczyk-jcp-146-124125-2017}.
PNP is a mean field method that does not consider the particles as individual entities, but investigates the concentration profiles as the probability of finding a center of  a particle in a certain location. 
The concentration  depends on the interaction energy of the ion with the average (mean) electrical potential, $\Phi(\mathbf{r})$, produced by all the charges in the system, including all the ions. 
The electrochemical potential in PNP  reads as
\begin{equation}
\mu_{i}^{\mathrm{PNP}}(\mathbf{r}) = \mu_{i}^{0} + kT\ln c_{i}(\mathbf{r}) + q_{i}\Phi(\mathbf{r}) .
\label{eq:elchempotpnp}
\end{equation} 
The excess term describing to two- and many-body correlations between ions and hard-sphere exclusion, therefore, is absent.
These correlations are sampled naturally in LEMC.

The concentration profiles are related to the mean electrical potential through Poisson's equation, namely
\begin{equation}
\nabla^{2} \Phi (\mathbf{r}) = - \dfrac{1}{\epsilon_0 \epsilon} \sum_{i}q_i c_i(\mathbf{r}) .
\label{eq:poisson}
\end{equation} 
The above equations are valid if we measure concentration in m$^{-3}$ (number density).
The results, however, will be shown in unit mol/dm$^{3}$ in order to make it easier to relate to usual concentration units.

The solution domain is different in the case of PNP and NP+LEMC.
In NP+LEMC, the interior of the membrane is part of the simulation cell, where electric field lines can protrude (it is an empty continuum with dielectric constant $\epsilon$).
In PNP, this region is excluded from the solution domain.
Neumann boundary conditions are prescribed on the surface of the membrane as detailed in our previous study.
On the nanopore's wall, we prescribe Neumann boundary condition that produces the desired surface charge: $\partial \Phi (\mathbf{r})/\partial \mathbf{n}_{\mathrm{W}}=\sigma_{\mathrm{pore}}(z)$, where $\sigma_{\mathrm{pore}}(z)$ is the prescribed surface charge at coordinate $z$ ($\sigma_{\mathrm{n}}$ or $\sigma_{\mathrm{x}}$) and $\mathbf{n}_{\mathrm{W}}$ is the normal vector on the surface of the pore wall.
The boundary condition $\mathbf{j}_{i}\cdot \mathbf{n}=0$ for the impenetrable membrane surface is also set.

On the membrane's surfaces that are perpendicular to the $z$-axis at $z= \pm H_{\mathrm{pore}}/2$ we impose the boundary conditions $\mathbf{j}_{i}(\mathbf{r})\cdot \mathbf{n}_{\mathrm{M}}=0$ and ${\partial \Phi (\mathbf{r})}/{\partial \mathbf{n}_{\mathrm{M}}}  = {\partial \Phi^{\mathrm{app}} (\mathbf{r})}/{\partial \mathbf{n}_{\mathrm{M}}}$
where $\mathbf{n}_{\mathrm{M}}$ is the outer normal and $\Phi^{\mathrm{app}}$ is the applied field used in the LEMC model. 
This solution mimics the LEMC case where there is an electric field across the membrane.

In the case of the two half-cylinders confining the solution domain, the same boundary conditions are prescribed as in the case of the NP+LEMC model.
Bulk concentrations $c_{i}^{\mathrm{L}}$ and $c_{i}^{\mathrm{R}}$ are set on the left and right hand side cylinders.
Dirichlet boundary conditions for the electrical potential are set: $\Phi^{\mathrm{L}}$ and $\Phi^{\mathrm{R}}$.
In practice, $\Phi^{\mathrm{L}}=0$ (left side is grounded) and $\Phi^{\mathrm{R}}=U$ ($U$ is the voltage).

To  solve the two-dimensional PNP system we use the Scharfetter--Gummel scheme which is based on a transformed formulation of the system in entropy  variables \cite{gummel1964self}. 
We use a two-dimensional finite element method for the actual implementation and a triangular mesh containing $100-200$ thousand elements.
The mesh is  non-uniform in order to obtain high accuracy, especially close to the pore entrances and  charged pore walls.

\section{Results}
\label{sec:results}

This paper studies the quantitative effect of changing the charge pattern (the values of $\sigma_{\mathrm{n}}$, $\sigma_{\mathrm{x}}$, $H_{\mathrm{n}}$, and $H_{\mathrm{x}}$) on the nanopore's wall.
We introduce special cases that we denote by strings ``$-+-$'', ``$-\,0\,-$'', ``$-\, -\, -$'' and so on as introduced earlier.
Some of these patterns are defined as ON states of the transistor (``$-\,0\,-$'' and ``$-\, -\, -$''), while ``$-+-$'' is defined as the OFF state.
This way, we can define a switch whose device function is the ratio of currents in the ON and OFF states, $I_{\mathrm{ON}}/I_{\mathrm{OFF}}$.
The larger this number is, the better the device works as a switch.

In this work, we use a 1:1 electrolyte with the same ionic diamaters for the cation and the anion ($0.3$ nm).
This choice makes a more straightforward comparison with PNP that cannot distinguish between ions of different sizes.
The dielectric constant is $\epsilon=78.5$, the temperature is $T=298.15$ K.
The bulk diffusion constant of both ion species is $1.334\cdot10^{-9}$ m$^{2}$/s, while the value inside the pore is ten times smaller \cite{matejczyk-jcp-146-124125-2017,madai-jcp-147-244702-2017}, a choice that does not qualitatively affect our conclusions.

In the case of 0.1 M concentration, this corresponds to about 800 ions in the LEMC simulations. 
An NP+LEMC calculation contained 80 iterations with LEMC simulations sampling 30 million configurations in an iteration.
Such a simulation lasted about 3 days.
This resulted in small statistical uncertainties for the currents; the error bars are smaller than the symbols with which the current data are plotted in the figures.
The PNP calculations, on the other hand, took only a few minutes.  

In this work, we will show cross-averaged axial concentration profiles computed as 
\begin{equation}
c_{i}(z)  =  \dfrac{1}{A_{\mathrm{eff}}} \int_{0}^{R_{\mathrm{eff}}} c_{i}(z,r)  2\pi r\, dr ,
\end{equation}
where $A_{\mathrm{eff}}=R_{\mathrm{eff}}^{2}\pi$ is the effective cross section of the pore that is accessible to the centers of the ions. 
For the hard sphere ions in LEMC $R_{\mathrm{eff}}=R_{\mathrm{pore}}-R_{i}$, while for the point ions in PNP $R_{\mathrm{eff}}=R_{\mathrm{pore}}$.
This definition makes profiles more comparable between LEMC and PNP.

\subsection*{Effect of charge pattern: changing surface charges  }
\label{subsec:chargep-sig}

As a first step, we vary the charge densities $\sigma_{\mathrm{x}}$ and $\sigma_{\mathrm{n}}$ and examine the resulting effect on the ionic current through the nanopore for a fixed geometry ($H_{\mathrm{n}}=3.4$ nm and $H_{\mathrm{x}}=3.2$ nm).
This current is driven by voltage $200$ mV; the concentration of the electrolyte is $c=0.1$ M on both sides of the membrane. 
These parameters are valid for all figures unless otherwise stated.

\begin{figure}[t!]
	\begin{center}
		\scalebox{0.67}{\includegraphics*{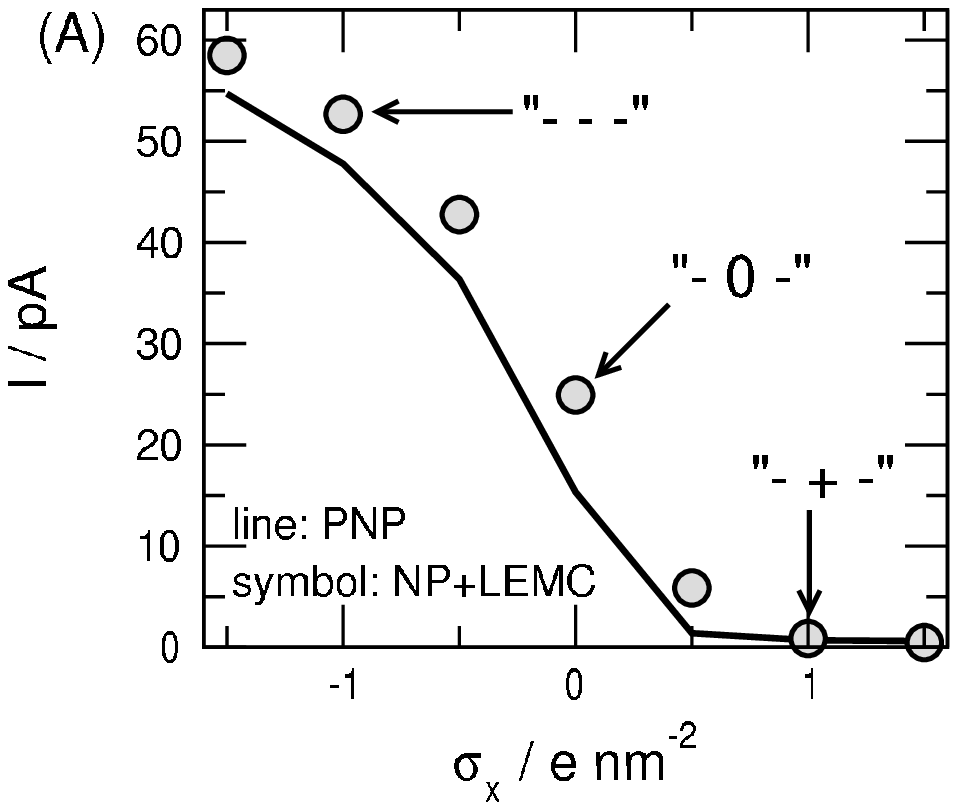}}\\ \vspace{0.5cm}
		\scalebox{0.7}{\includegraphics*{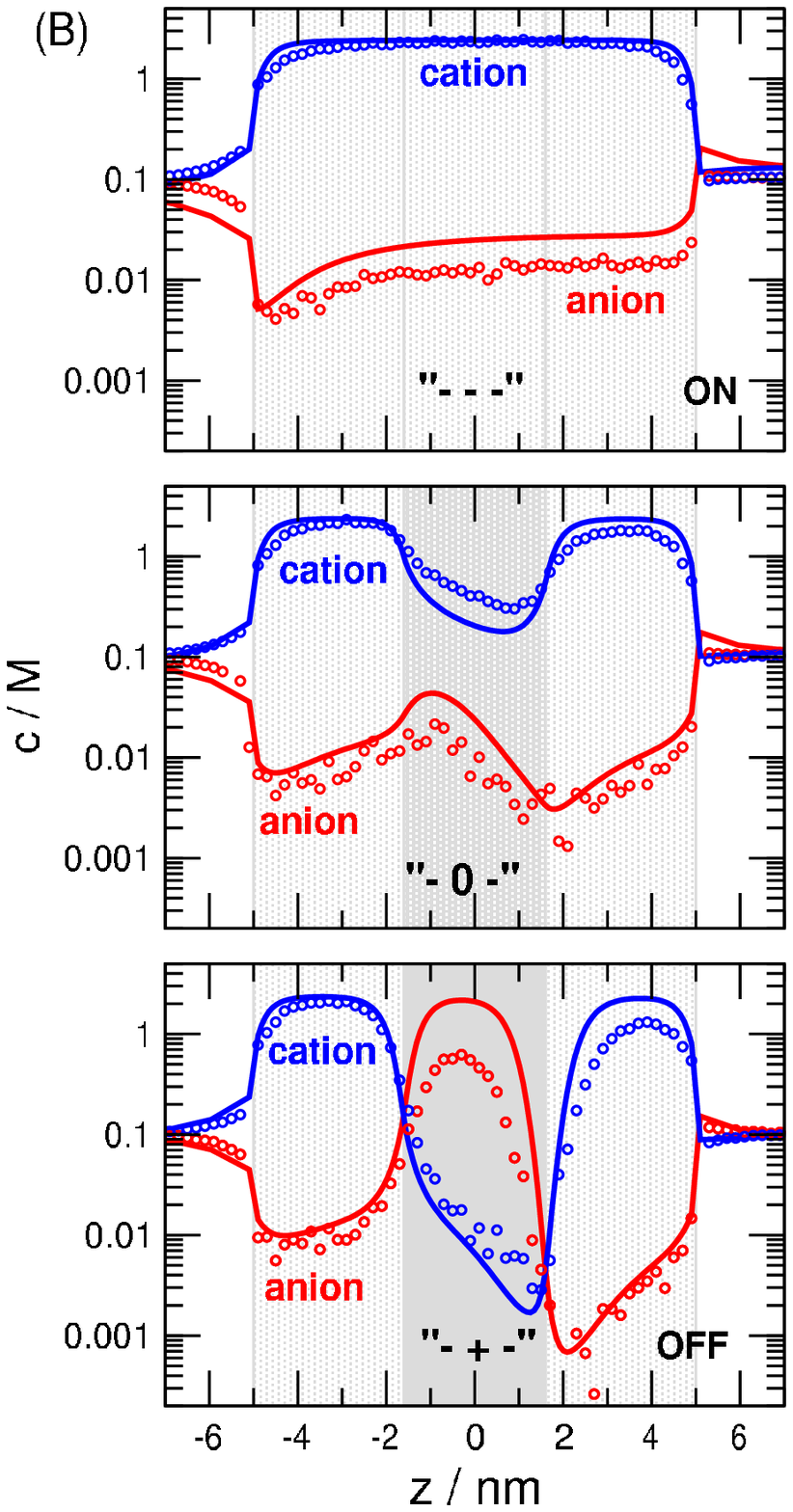}}
	\end{center}
	\caption{
		(A) Current as a function of $\sigma_{\mathrm{x}}$ while $\sigma_{\mathrm{n}}=-1$ $e/\mathrm{nm}^{2}$ is kept fixed. 
		Selected charge patterns are indicated with ``$- - -$'' (ON), ``$-\,0\, -$'', and ``$-+-$'' (OFF). 
		Increasing $\sigma_{\mathrm{x}}$ makes the x region more positive, so the $I(-\sigma_{\mathrm{x}})$ function is monotonically decreasing.
		(B) Concentration profiles for these selected charge patterns.  
		Widths of the regions are $H_{\mathrm{x}}=3.2$ and $H_{\mathrm{n}}=3.4$ nm, electrolyte concentration is $c=0.1$ M, voltage is 200 mV.
		Symbols and lines denote NP+LEMC and PNP results, respectively, here and in all the remaining figures unless otherwise stated.
	}
	\label{Fig2}
\end{figure} 

\afterpage{\clearpage}

Figure \ref{Fig2}A shows results for a fixed $\sigma_{\mathrm{n}}=-1$ $e/\mathrm{nm}^{2}$ and varying $\sigma_{\mathrm{x}}$.
The negative value of $\sigma_{\mathrm{n}}$ makes the nanopore cation-selective due to the large surface charge and small pore radius.
Therefore, the main charge carrier is the cation.
The current of the anion remains below 0.5 pA.
The anions have depletion zones in the two n regions as seen in Fig.\ \ref{Fig2}B.
Whether the anions have depletion zones in the central x region depends on the value of $\sigma_{\mathrm{x}}$, but this is irrelevant, because they already have depletion zones in the n regions.

In this model, the value of $\sigma_{\mathrm{x}}$ tunes the depletion zones of the cations, and, thus, the cation current.
In the case of $\sigma_{\mathrm{x}}=-1$ $e/\mathrm{nm}^{2}$ (``$- - -$''), cations do not have a depletion zone in the middle, so they carry electrical current. 
This is an ON state of the device (top panel of Fig.\ \ref{Fig2}B).
Increasing $\sigma_{\mathrm{x}}$ towards positive values, the depletion zone of cations gradually appears (see Fig.\ \ref{Fig2}B) and the cation current gradually decreases (see Fig.\ \ref{Fig2}A). 

\subsection*{Effect of charge pattern: changing region widths}
\label{subsec:chargep-H}

\begin{figure}[t]
	\begin{center}
		\scalebox{0.55}{\includegraphics*{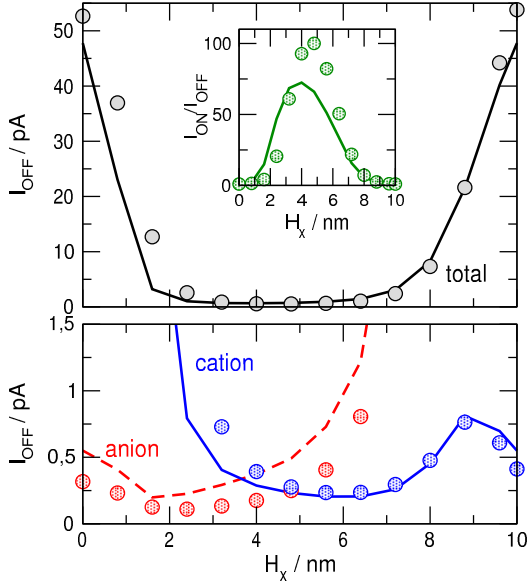}}
	\end{center}
	\caption{Currents in the OFF state (``$-+-$'') through nanopores with varying region lengths. 
		The total length, $H_{\mathrm{pore}}=2H_{\mathrm{n}}+H_{\mathrm{x}}=10$ nm, is kept fixed. 
		The results are shown as functions of $H_{\mathrm{x}}$. 
		Top panel shows the total current, while the bottom panel shows the cation and anion currents. 
		The inset of the top panel shows the $I_{\mathrm{ON}}/I_{\mathrm{OFF}}$ ratio, where the charge pattern of the ON state is ``$---$'' (its current is independent of $H_{\mathrm{x}}$). 
	}
	\label{Fig3}
\end{figure}

Next, we fix the charge densities and change the geometry, namely, the widths $H_{\mathrm{x}}$ and $H_{\mathrm{n}}$ for a fixed pore radius.
Particularly, we examined the effect of changing the relative widths of the x and n regions while keeping the total width $H_{\mathrm{pore}}=2H_{\mathrm{n}}+H_{\mathrm{x}}=10$ nm fixed.
In the ON state (``$-\, -\, -$''), there is no difference between these regions, so we need to examine the OFF state (``$-+-$'') only.
We plot the currents in the OFF state as functions of $H_{\mathrm{x}}$ in Fig.\ \ref{Fig3}.

The top panel showing the total current exhibits a minimum that is better observed in the inset that shows $I_{\mathrm{ON}}/I_{\mathrm{OFF}}$.
Because $I_{\mathrm{ON}}$ does not depend on $H_{\mathrm{x}}$, the ratio is proportional to the reciprocal of $I_{\mathrm{OFF}}$.
The minimum in $I_{\mathrm{OFF}}$, therefore, corresponds to a maximum in the ratio characterizing the quality of the device as a switch.

The explanation of this extremum can be depicted from the bottom panel of Fig.\ \ref{Fig3}.
For small $H_{\mathrm{x}}$ values, the pore is largely negatively charged, so the main charge carrier is the cation.
For large $H_{\mathrm{x}}$ values, the situation is reversed: the main charge carrier is the anion.
The minimum of the current occurs at a $H_{\mathrm{x}}$ value, where both regions have sufficient size to produce sufficiently deep depletion zones for both ionic species: for cations in the n regions, while for anions in the x region.
This value is somewhere around $H_{\mathrm{x}}=5$ nm.

\begin{figure}[t]
	\begin{center}
		\scalebox{0.55}{\includegraphics*{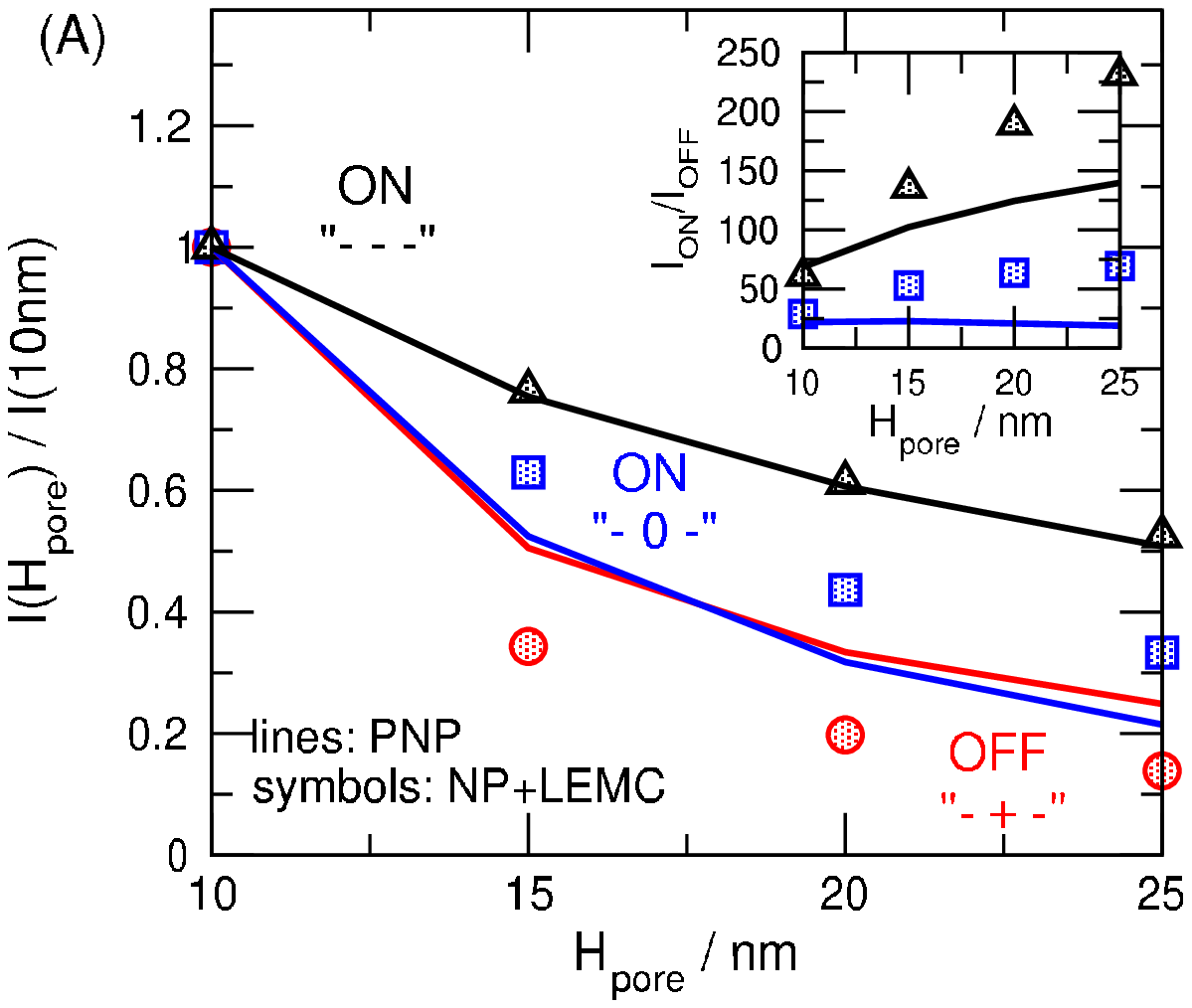}}\\ \vspace{0.3cm}
		\scalebox{0.55}{\includegraphics*{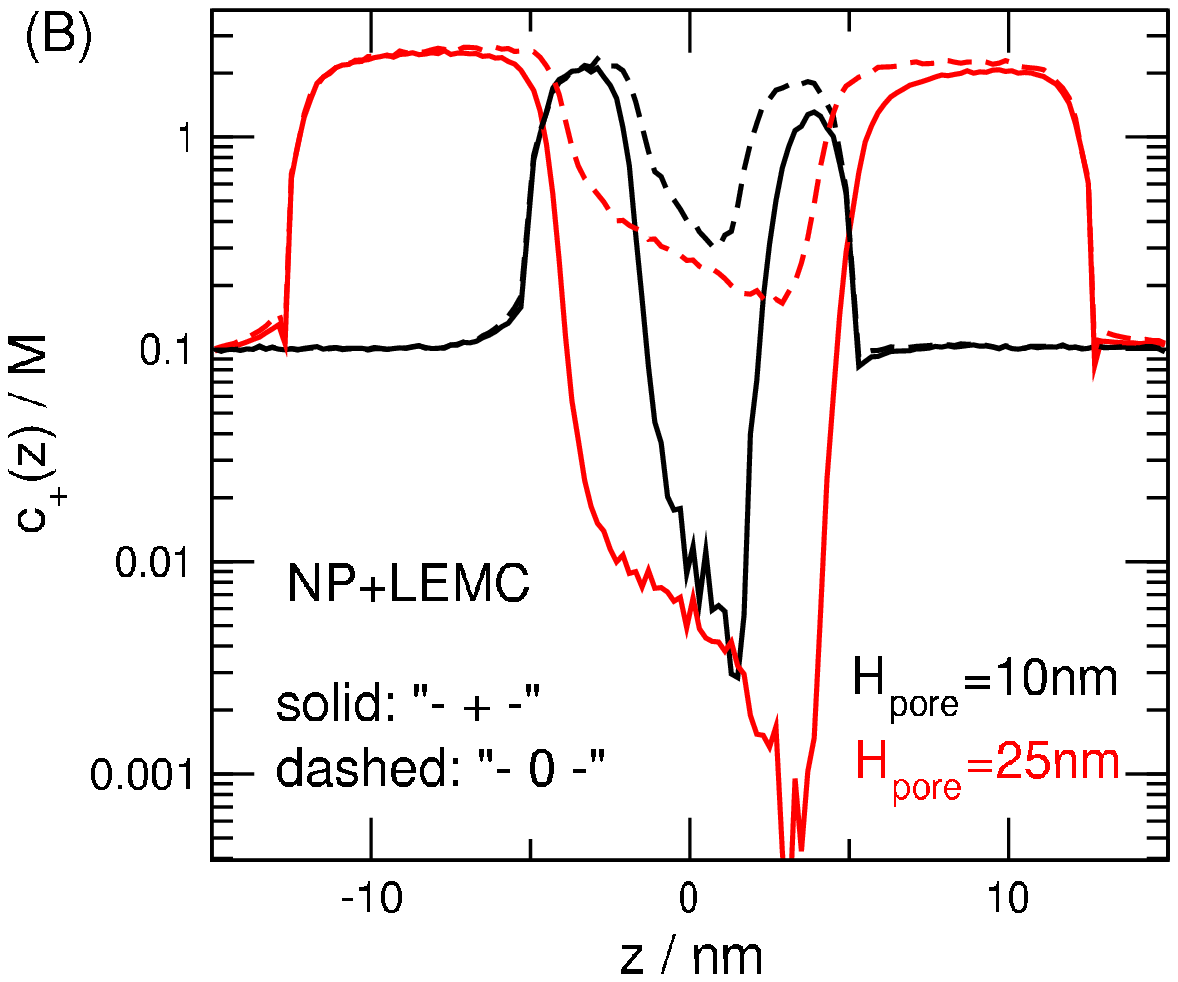}}
	\end{center}
	\caption{(A) 
		Total currents as functions of pore length, $H_{\mathrm{pore}}$, for various charge patterns with $H_{\mathrm{n}}/H_{\mathrm{x}}=1.0625$ kept fixed. 
		The currents are normalized with the values at $H_{\mathrm{pore}}=10$ nm.
		The inset shows the $I_{\mathrm{ON}}/I_{\mathrm{OFF}}$ ratio for the two cases where the ON states are defined either with ``$-\,0\,-$'' or ``$-\, -\, -$''.
		(B) Concentration profiles of the anions (the charge carriers) for $H_{\mathrm{pore}}=10$ nm (black) and $H_{\mathrm{pore}}=25$ nm (red) for charge patterns ``$-+-$'' (solid) and ``$-\,0\,-$'' (dashed) as obtained from NP+LEMC simulations. 
	}
	\label{Fig4}
\end{figure}

\subsection*{Effect of pore length}
\label{subsec:poresize}

Figure \ref{Fig4} shows the result for the case, where the $H_{\mathrm{n}}/H_{\mathrm{x}}$ ratio is kept fixed (at the value of 1.0625) and the total pore length, $H_{\mathrm{pore}}=2H_{\mathrm{n}}+H_{\mathrm{x}}$ is changed.
Figure \ref{Fig4}A shows the relative currents for the ``$-\, -\, -$'', ``$-\,0\,-$'' (ON), and ``$-+-$'' (OFF) states.
We plot relative currents (normalized by the values at $H_{\mathrm{pore}}=10$ nm) because we are rather interested in how fast the currents decrease as functions of $H_{\mathrm{pore}}$ in the different cases (ON and OFF).

Figure \ref{Fig4}A shows that the currents decrease faster in the OFF state than in the ON states.
This results in an increasing $I_{\mathrm{ON}}/I_{\mathrm{OFF}}$ ratio as shown in the inset of Fig.\ \ref{Fig4}A.
The explanation is the deepening depletion zones with increasing $H_{\mathrm{pore}}$ (Fig.\ \ref{Fig4}B).

The inset of Fig.\ \ref{Fig4}A also shows that the ON/OFF ratio exhibits a saturation behavior so we can extrapolate to large $H_{\mathrm{pore}}$ values that are more common in experiments, but harder to attain with particle simulations such as LEMC.
Summarized, increasing pore length promotes the formation of depletion zones due to weakening electrostatic correlations between neighboring zones.

\subsection*{Effect of pore radius and concentration}
\label{subsec:poreradius}

We discuss the effect of nanopore radius and concentration together, because concentration determines $\lambda_{\mathrm{D}}$ (see Eq.\ \ref{eq:lambdaD}), so $R_{\mathrm{pore}}$ and $c$ influence the $R_{\mathrm{pore}}/\lambda_{\mathrm{D}}$ ratio that distinguishes nanopores from micropores as discussed in the Introduction.
In this work, we study the effect of changing $R_{\mathrm{pore}}/\lambda_{\mathrm{D}}$ in three ways.
First, we keep $\lambda_{\mathrm{D}}$ constant by fixing the concentration at $c=0.1$ M and vary $R_{\mathrm{pore}}$, then we do the reverse.
Finally, we change both $R_{\mathrm{pore}}$ and $c$ while keeping $R_{\mathrm{pore}}/\lambda_{\mathrm{D}}$ fixed. 

\begin{figure}[t!]
	\begin{center}
		\scalebox{0.55}{\includegraphics*{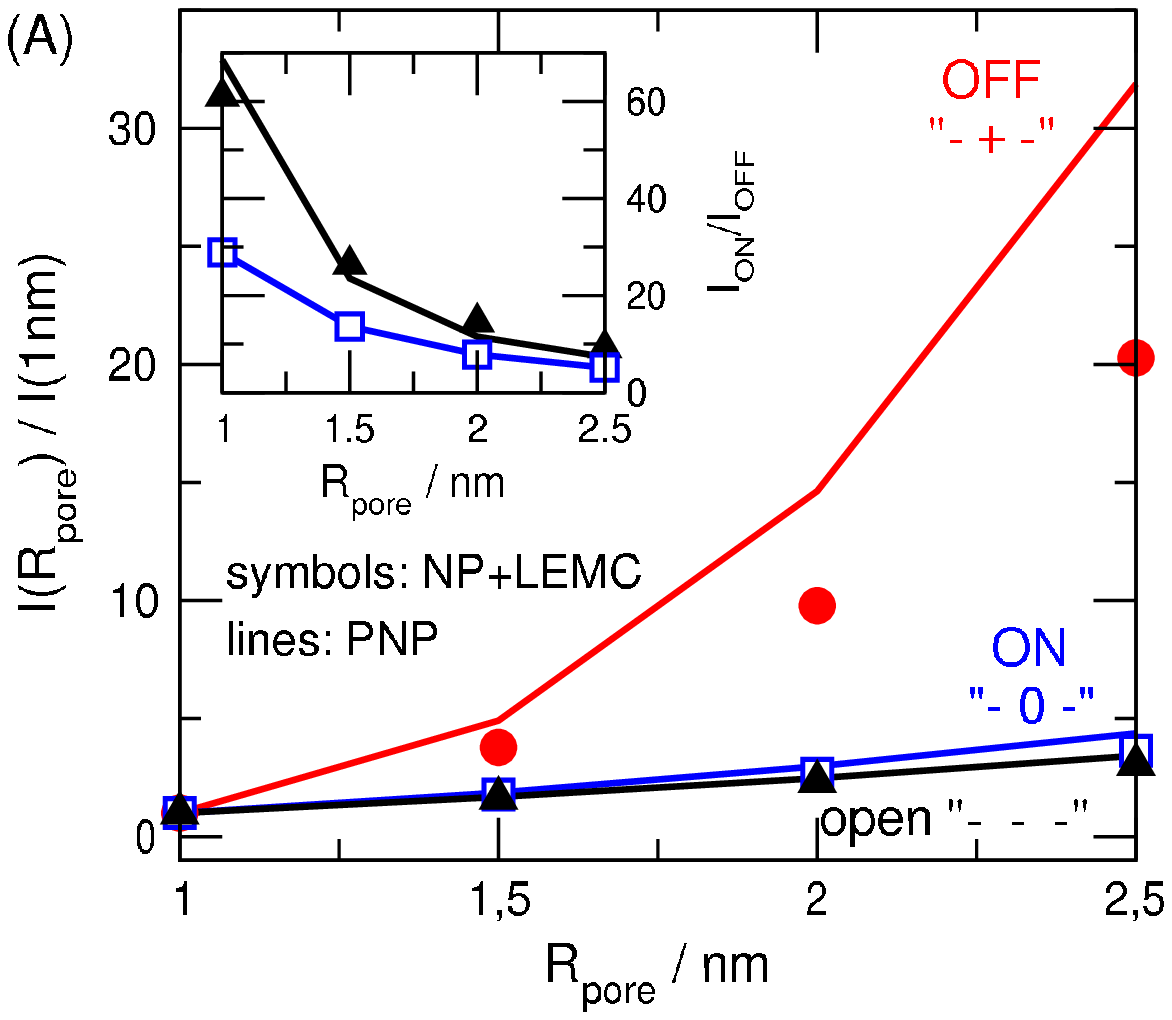}}\\ \vspace{0.3cm}
		\scalebox{0.55}{\includegraphics*{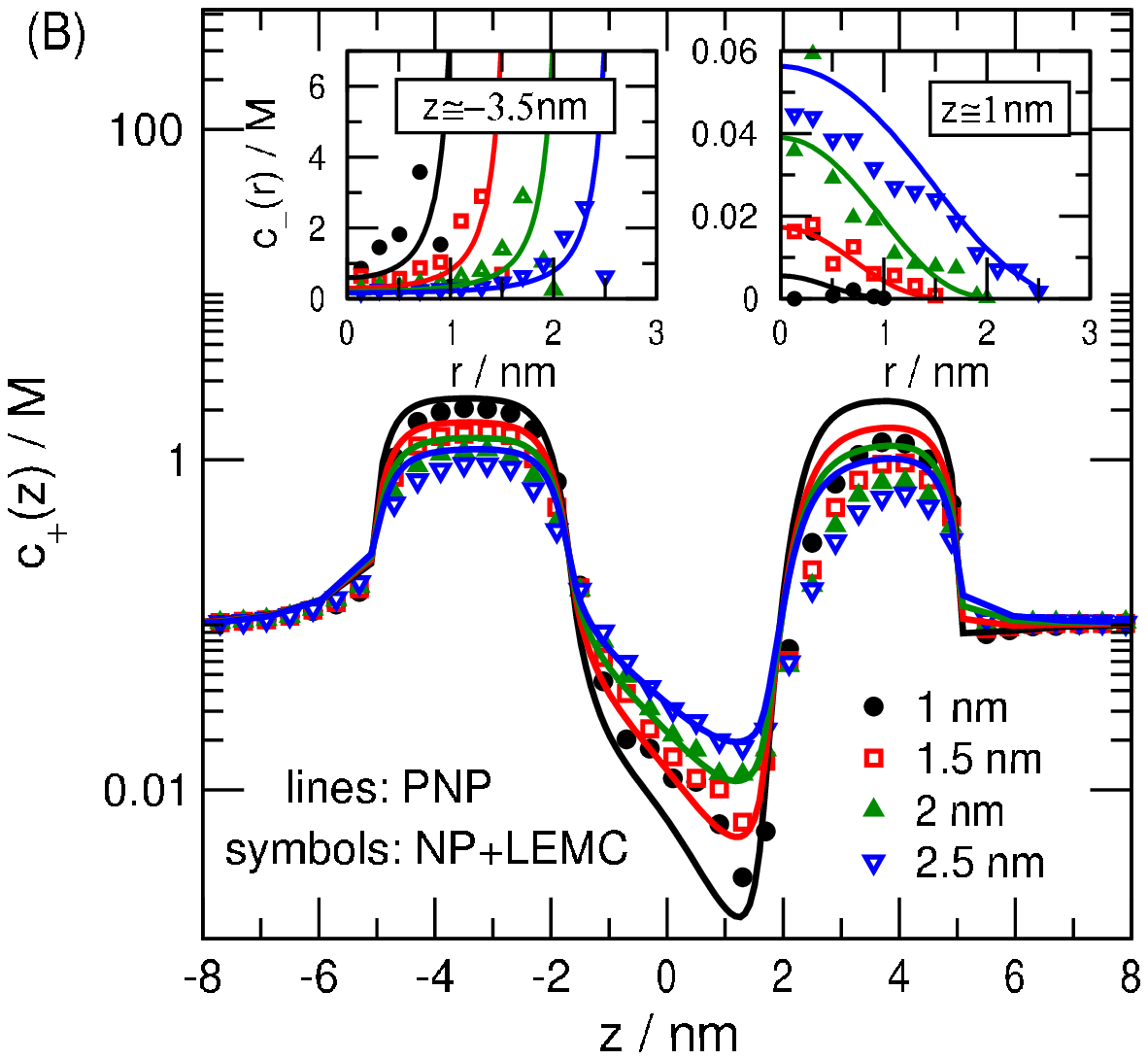}}
	\end{center}
	\caption{
		(A) Total currents as functions of pore radius, $R_{\mathrm{pore}}$, for various charge patterns for  $H_{\mathrm{n}}/H_{\mathrm{x}}=1.0625$ and $H_{\mathrm{pore}}=10$ nm. 
		The currents are normalized with the values at $R_{\mathrm{pore}}=1$ nm.
		The inset shows the $I_{\mathrm{ON}}/I_{\mathrm{OFF}}$ ratio for the two cases where the ON states are defined either with ``$-\,0\,-$'' or ``$-\, -\, -$''.
		(B) Axial concentration profiles of the cations (the charge carriers) for various $R_{\mathrm{pore}}$ values for charge pattern ``$-+-$'' (OFF state). 
		The insets show the radial concentration profiles for $z\approx -3.5$ nm (at a peak) and $z\approx 1$ nm (at the depletion zone).}
	\label{Fig5}
\end{figure}

Figure \ref{Fig5}A shows the normalized currents as functions of $R_{\mathrm{pore}}$ for the OFF (``$-+-$'') and the two ON (``$-\,0\,-$'' and ``$-\, -\, -$'') cases.
Here, we normalize with the currents at $R_{\mathrm{pore}}=1$ nm.
The relative current in the OFF state decreases faster with decreasing $R_{\mathrm{pore}}$ than in the OFF state, which, in turn, results in increasing $I_{\mathrm{ON}}/I_{\mathrm{OFF}}$ ratios with decreasing $R_{\mathrm{pore}}$ as shown by the inset.

Figure \ref{Fig5}B shows the cross-section-averaged axial concentration profiles of the cations, $c_{+}(z)$, in the OFF state for different pore radii.
As $R_{\mathrm{pore}}$ decreases, the depletion zones in the middle get deeper, so the current decreases as Fig.\ \ref{Fig5}A shows. 
The two insets show the radial concentration profiles, $c_{+}(r)$, at two characteristic axial positions: $z\approx -3.5$ nm is a peak, while $z\approx 1$ nm is a depletion zone.
The profiles at  $z\approx -3.5$ nm show that the cations are attracted to the pore wall and their concentrations decline approaching the pore centerline ($r\sim 0$).
The absence of a bulk electrolyte along the centerline is more apparent from the  $c_{+}(r)$ profiles for $z\approx 1$ nm showing that concentrations never reach the bulk value (0.1 M).

\begin{figure}[t]
	\begin{center}
		\scalebox{0.55}{\includegraphics*{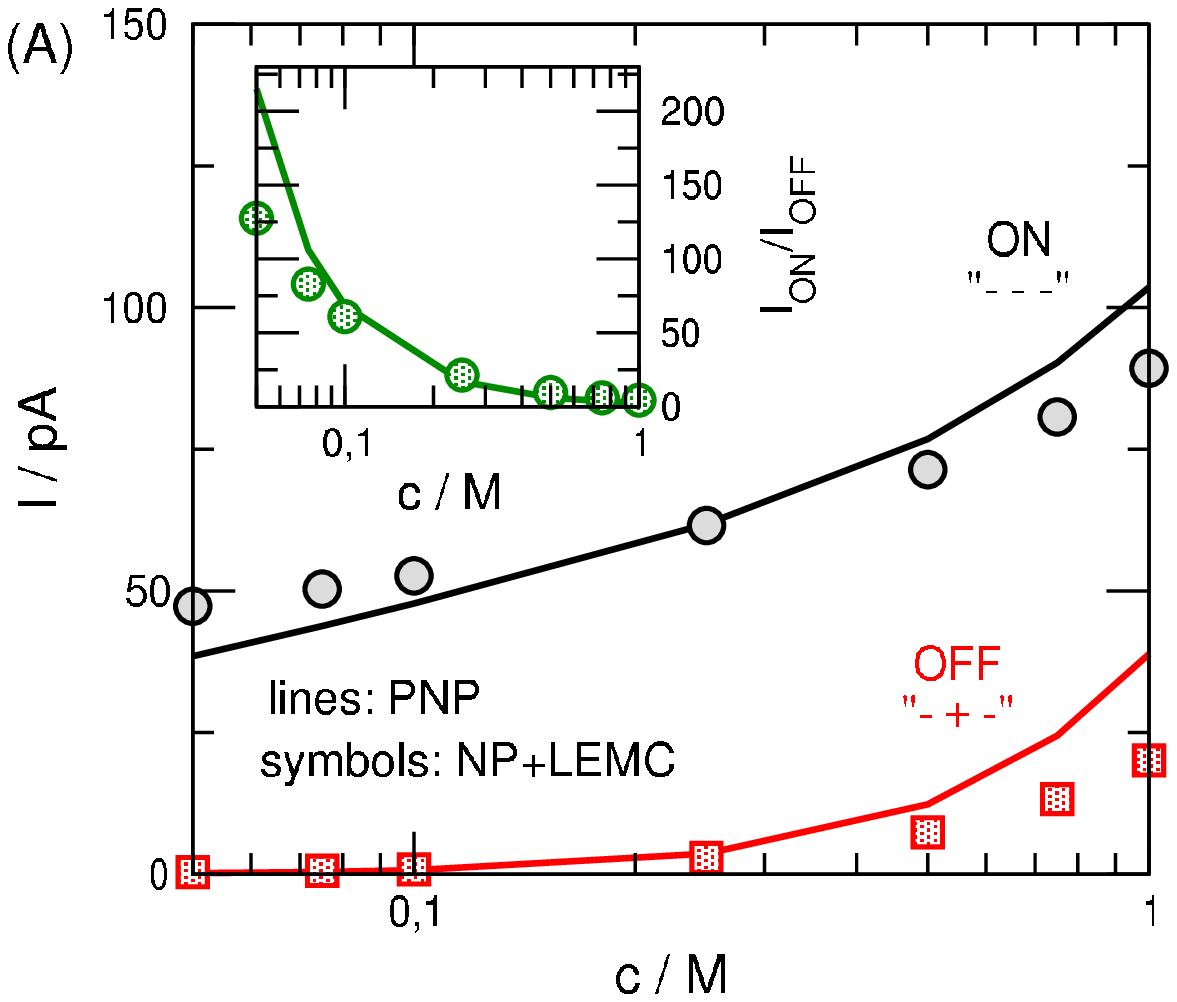}}\\ \vspace{0.3cm}
		\scalebox{0.55}{\includegraphics*{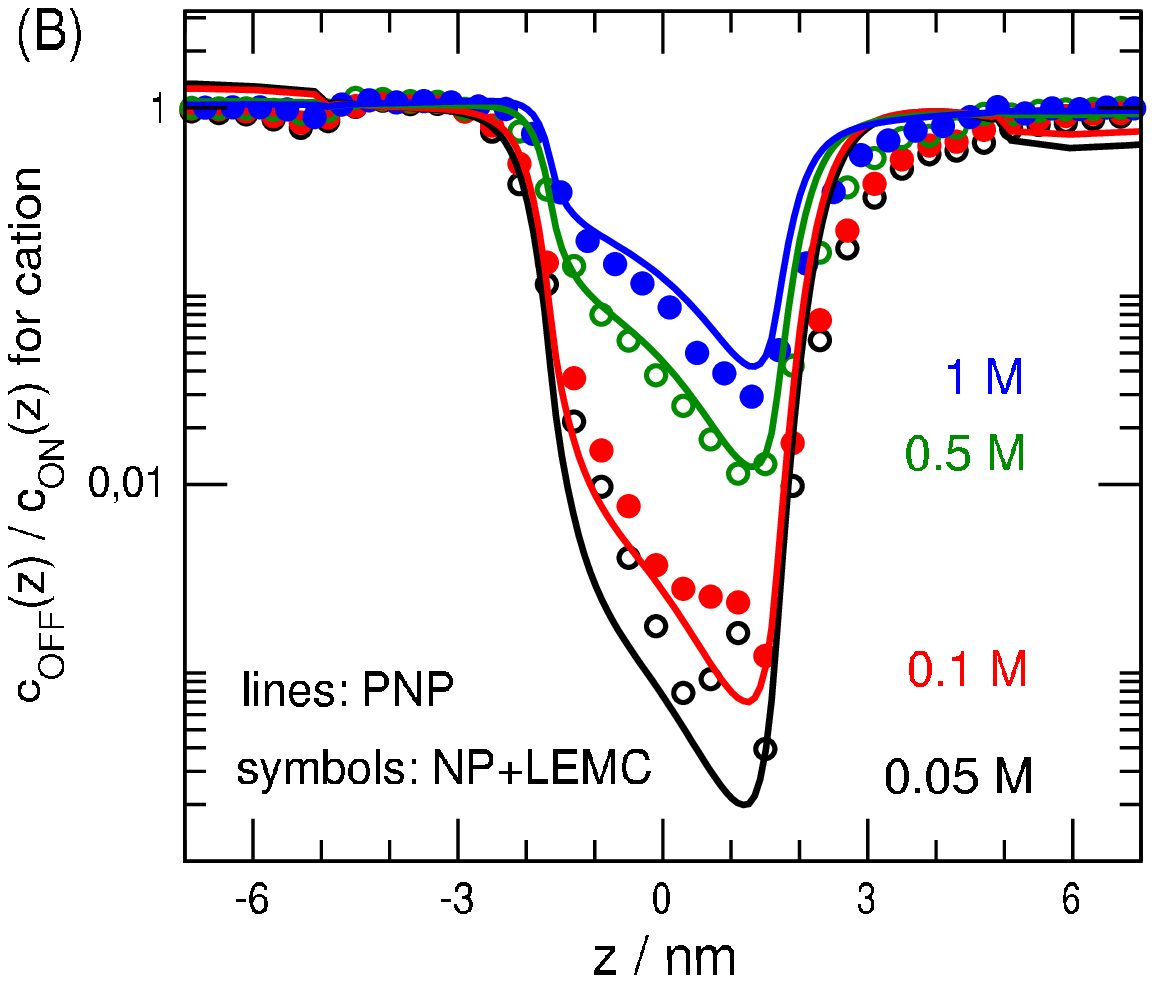}}
	\end{center}
	\caption{
		(A) Concentration dependence of the current in the ON (``$-\, -\, -$'') and OFF (``$-+-$'') states. 
		The inset shows the $I_{\mathrm{ON}}/I_{\mathrm{OFF}}$ ratio.
		(B) Ratio of cation concentration profiles in the OFF and ON states for different bulk concentrations.}
	\label{Fig6}
\end{figure}

Next, we study the effect of changing $R_{\mathrm{pore}}/\lambda_{\mathrm{D}}$ by keeping $R_{\mathrm{pore}}$ fixed at 1 nm and changing $\lambda_{\mathrm{D}}$ through varying concentration from $c=0.05$ M to $c=1$ M (it corresponds to changing the Debye length from $\lambda_{\mathrm{D}}=1.36$ nm to $\lambda_{\mathrm{D}}=0.304$ nm).
Figure \ref{Fig6}A shows the currents for the ``$-\, -\, -$'' (ON) and ``$-+-$'' (OFF) states.
Both currents decrease with decreasing concentration, but the OFF-state current decreases faster than the ON-state current.
This results in a increasing $I_{\mathrm{ON}}/I_{\mathrm{OFF}}$ ratio with decreasing $c$ (see inset).
The explanation again follows from the behavior of depletion zones.

Figure \ref{Fig6}B shows the cation concentration profiles in the OFF state divided by the profiles in the ON state.  
The behavior of these curves for different bulk concentration reveals that the cations have deeper depletion zones compared to the ON state for smaller bulk concentrations.
Because the $c_{\mathrm{ON}}(z)/c_{\mathrm{OFF}}(z)$ ratio is a first-order determinant of the $I_{\mathrm{ON}}/I_{\mathrm{OFF}}$ ratio, this ratio increases with decreasing $c$ due to deepening depletion zones in the OFF state relative to the ON state at same $c$.

\begin{figure}[t!]
	\begin{center}
		\scalebox{0.55}{\includegraphics*{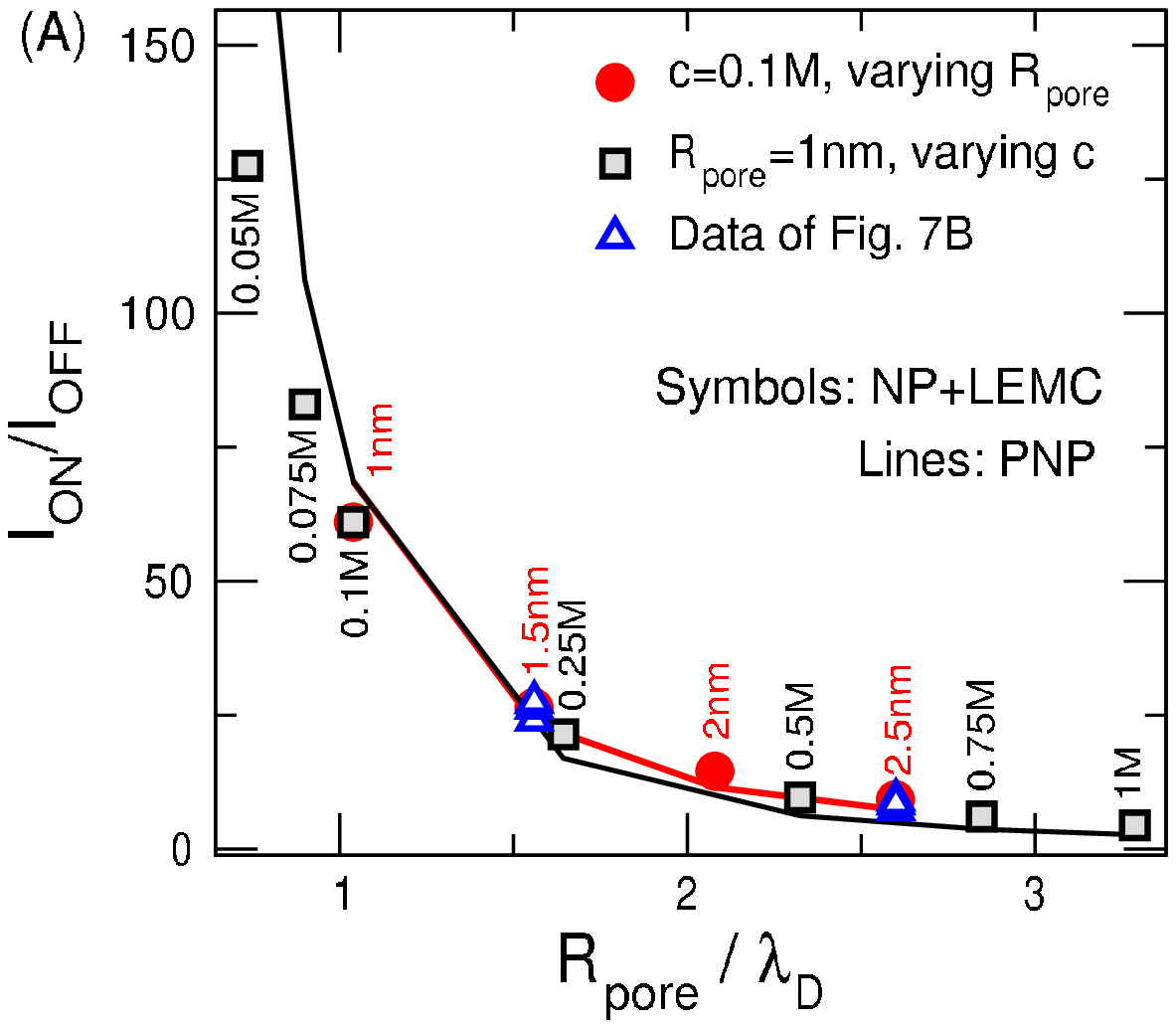}}\\ \vspace{0.1cm}
		\scalebox{0.55}{\includegraphics*{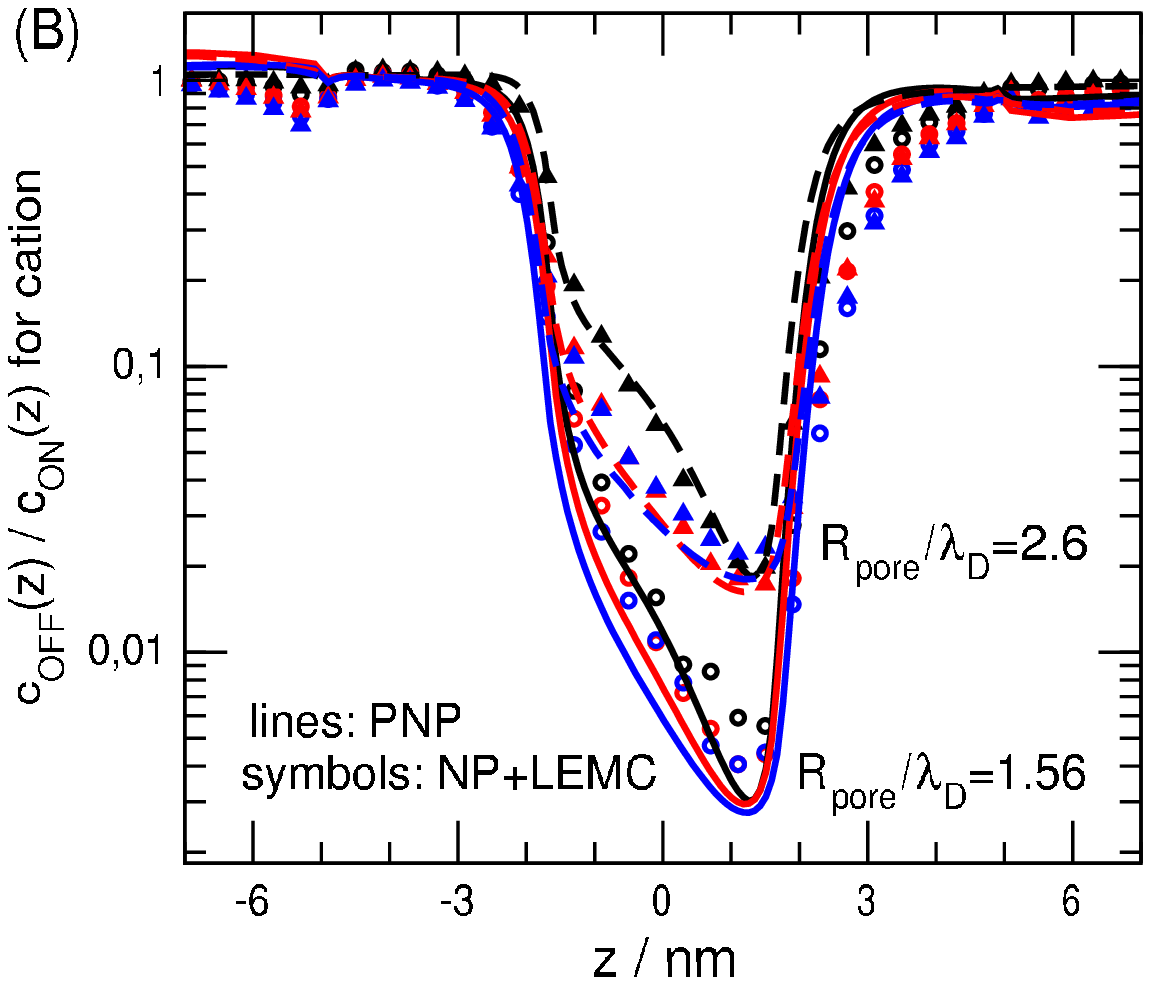}}
	\end{center}
	\caption{(A) The $I_{\mathrm{ON}}/I_{\mathrm{OFF}}$ ratio as a function of the $R_{\mathrm{pore}}/\lambda_{\mathrm{D}}$ variable for the cases, when we change $R_{\mathrm{pore}}$ at fixed $\lambda_{\mathrm{D}}$ ($c=0.1$ M, red), and when we change $\lambda_{\mathrm{D}}$ by changing concentration for a fixed $R_{\mathrm{pore}}=1$ nm (black).
		The numbers near symbols indicate pore radii (red) or concentration (black).
		(B) Ratio of cation concentration profiles in the OFF and ON states for combinations of $R_{\mathrm{pore}}$ and $\lambda_{\mathrm{D}}$ for fixed $R_{\mathrm{pore}}/\lambda_{\mathrm{D}}=1.56$ (solid lines and open symbols) and 2.6 (dashed lines and closed symbols) ratios. From bottom to top, the curves correspond to the following $(R_{\mathrm{pore}}/\mathrm{nm};\,c/\mathrm{M})$ pairs:
		$(1.924;\,0.0563)$ (blue), $(1.5;\,0.1)$ (red), $(1;\,0.225)$ (black) for $R_{\mathrm{pore}}/\lambda_{\mathrm{D}}=1.56$  
		and
		$(3.5;\,0.0511)$ (blue), $(2.5;\,0.1)$ (red), $(1;\,0.626)$ (black) for $R_{\mathrm{pore}}/\lambda_{\mathrm{D}}=2.6$.
		The $I_{\mathrm{ON}}/I_{\mathrm{OFF}}$ values for these points are indicated by blue triangles in Fig.\ \ref{Fig7}A.
	}
	\label{Fig7}
\end{figure} 

Finally, we performed simulations for two fixed values of  $R_{\mathrm{pore}}/\lambda_{\mathrm{D}}$ (1.56 and 2.6) by using various combinations of $R_{\mathrm{pore}}$ and $c$ (see caption of Fig.\ \ref{Fig7}).
These ways of studying $R_{\mathrm{pore}}/\lambda_{\mathrm{D}}$ dependence are summarized in Fig.\ \ref{Fig7}A by plotting the  $I_{\mathrm{ON}}/I_{\mathrm{OFF}}$ ratio against the $R_{\mathrm{pore}}/\lambda_{\mathrm{D}}$ ratio.
The fact that the data are located along a single curve shows a scaling behavior: we can either use a wide pore with small concentration (if fabrication of a narrow pore is the limiting factor), or a narrow pore with large concentration (if using small concentrations is the limiting factor due, for example, to detecting small currents).

Figure \ref{Fig7}B shows the $c_{\mathrm{OFF}}(z)/c_{\mathrm{ON}}(z)$ cation profiles for those combinations of $R_{\mathrm{pore}}$ and $\lambda_{\mathrm{D}}$ (changed via changing $c$) that provide the 1.56 and 2.6 values for the ratio. 
The coincidence of the curves shows that scaling is valid not only for current ratios, but also for concentration ratios.
Such scaling behavior is always advantageous in designing devices for a given response function.


\section{Discussion}
\label{sec:discussion}

\subsection*{Controlling with pH}
\label{subsec:pH}

Manipulating charge pattern on the nanopore surface is a non-trivial chemical treatment for which, generally, the nanopore needs to be removed from the measuring cell.
There is, however, a way of altering charge pattern during the measurement by changing the pH of the bath electrolytes in the measuring cell.
If there are different chemical groups on the pore surface in the x and n regions that respond differently to pH (protonation vs.\ deprotonation), their charge can be changed with varying pH.

For example, if the surfaces of the n and x regions are functionalized by carboxyl and amino groups, respectively, they become  negative and positive, respectively, at neutral pH (``$-+-$'', OFF state).
Changing the pH to acidic, the carboxly groups in the n regions get protonated and become neutral.
The amino groups of the x region, in the meantime, remain positive, so this results in a  ``$0\, +\,0$'' (ON) state.
Changing the pH to basic, the amino groups in the x region get deprotonated and become neutral.
The carboxyl groups of the n regions, in the meantime, remain negative, so this results in a ``$-\, 0\,-$'' (also ON) state.

\begin{figure}[t]
	\begin{center}
		\scalebox{0.55}{\includegraphics*{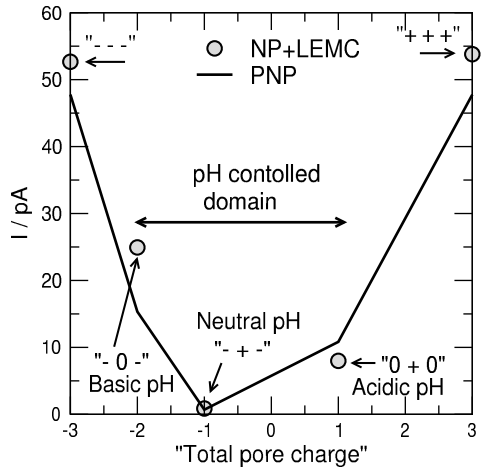}}
	\end{center}
	\caption{Demonstration of the effect of pH by plotting the current against the ``total pore charge'' characterizing the asymmetry of the pore's charge distribution.
		Assuming that the n and x regions have about equal lengths, this dimensionless number is obtained by  $\sum_{k=1}^{3}\sigma_{k}/\sigma_{0}$, where $\sigma_{k}$ is the surface charge of region $k$ and $\sigma_{0}=1$ $e/\mathrm{nm}^{2}$.
		OFF states of the transistor are present in cases when this number is close to zero, namely, when depletion zones for both ionic species are present (``$-+-$''). 
		For the example given in the main text (carboxyl and amino groups), this charge pattern is present at neutral pH.
		ON states are present when depletion zones for one of the ionic species are absent.
		The charge patterns ``$0\,+\,0$'' or ``$-\,0\,-$'' can be produced by tuning the pH towards acidic or basic, respectively.
	}
	\label{Fig8}
\end{figure}

The results are shown in Fig.\ \ref{Fig8}.
Currents are shown as functions of a quantity depicted  as ``total pore charge''.
This is practically the sum of the magnitudes (with sign) of surface charges in the three regions. 
This figure is closely related to Fig.\ \ref{Fig3}, where this ``total pore charge'' was controlled with $H_{\mathrm{x}}$.
There, the minimum of the curve was at $H_{\mathrm{x}}\approx 5$ nm, that corresponds to zero ``total pore charge''.
In that case, there are both positive and negative regions in a balanced ratio so that depletion zones of both cations and anions form in an optimal way so that current is minimized.
Here, the OFF state (``$+-+$'') appear at neutral pH, while the pore can be switched ON with changing pH in any direction \cite{kalman_am_2008}. 

\subsection*{Controlling with charge vs.\  potential}
\label{subsec:charge-pot}

\begin{figure}[t]
	\begin{center}
		\scalebox{0.55}{\includegraphics*{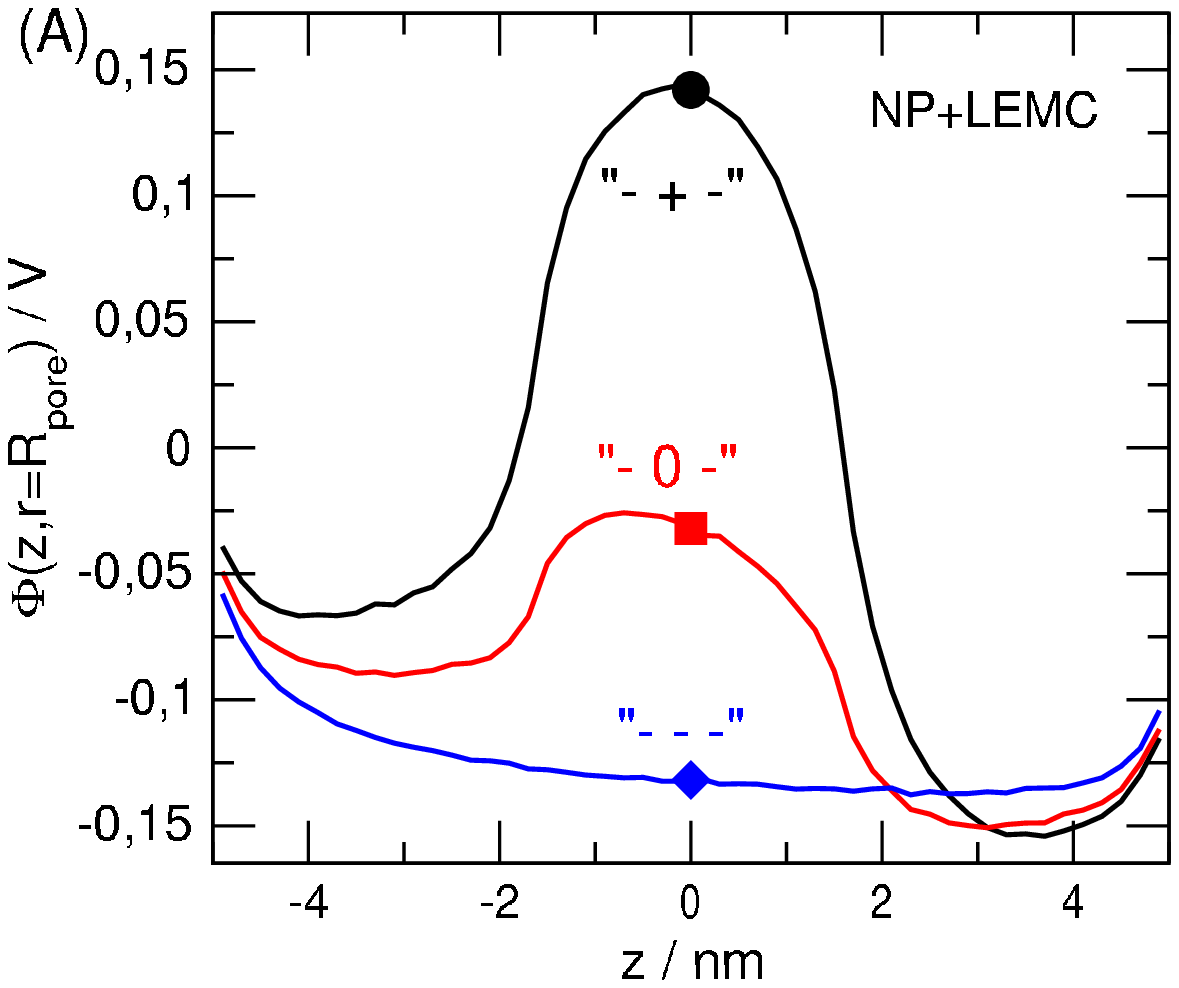}}\\ \vspace{0.1cm}
		\scalebox{0.55}{\includegraphics*{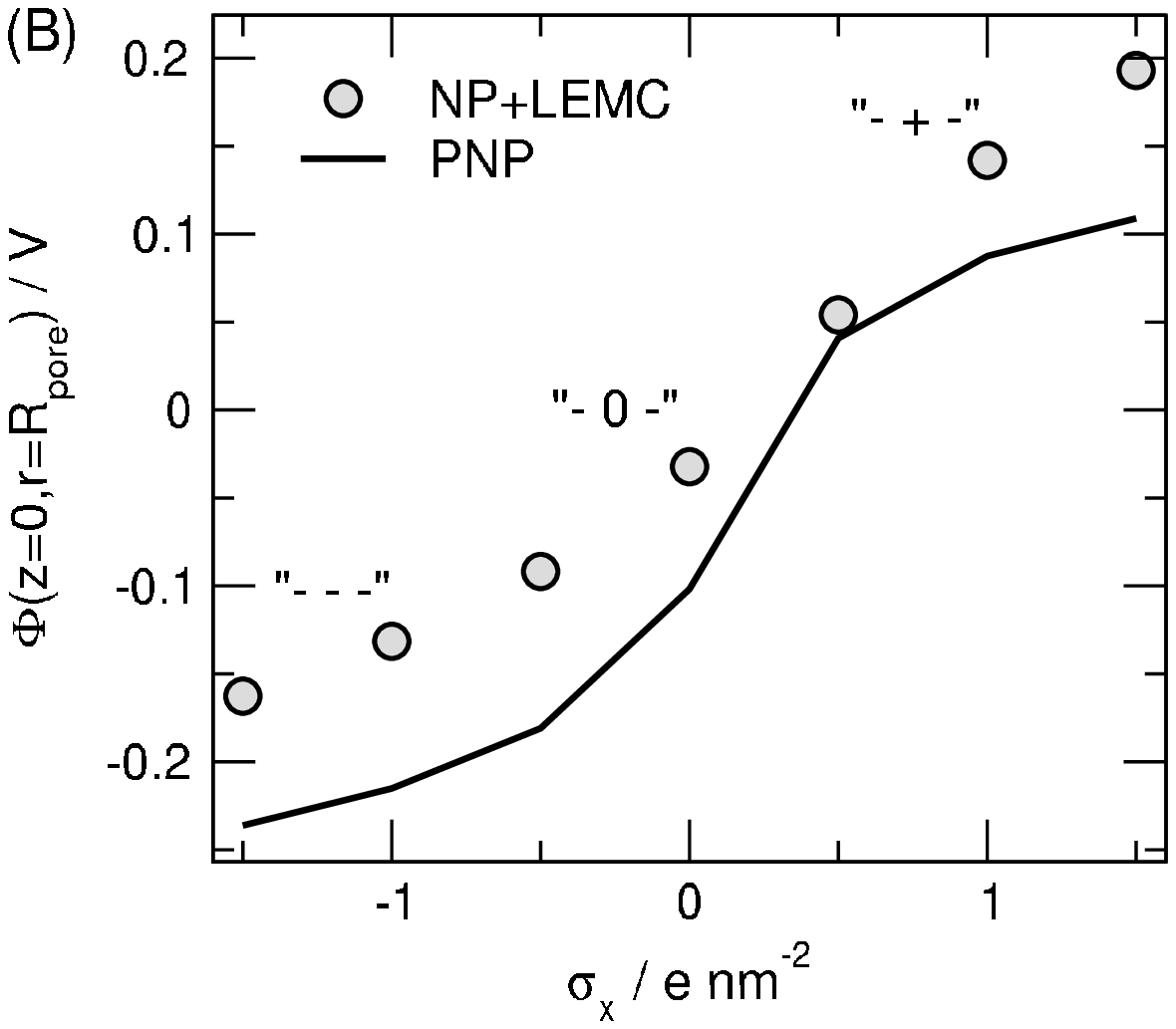}}
	\end{center}
	\caption{(A) The value of the mean electrical potential on the surface of the pore wall ($r=R_{\mathrm{pore}}$) for three selected charge patterns as obtained from NP+LEMC calculations.
		(B) The value of this potential in the center of the pore ($z=0$, $r=R_{\mathrm{pore}}$ shown with larger symbols in panel A) as a function of $\sigma_{\mathrm{x}}$.  
		The figure demonstrates the monotonic relation between surface charge density, $\sigma_{\mathrm{x}}$, and surface potential, $\Phi(z=0,r=R_{\mathrm{pore}})$.
	}
	\label{Fig9}
\end{figure}

Controlling surface charge is quite different from controlling the electrical potential from a practical point of view, but from a modeling point of view, they are similar because charge is always related to  electrical potential through Poisson's equation (Eq.\ \ref{eq:poisson}).
To show this, we plot the electrical potential profile on the surface of the nanopore, $r=R_{\mathrm{pore}}$, for different values of $\sigma_{\mathrm{x}}$ in Fig.\ \ref{Fig9}A. 
The potential profile changes in zone x, because it is not an imposed quantity.
The magnitude of the potential characterized by its value in the center, $z=0$, depends unambiguously on $\sigma_{\mathrm{x}}$.
As Fig.\ \ref{Fig9}B shows, there is a monotonic relation between charge, $\sigma_{\mathrm{x}}$, and potential, $\Phi(0,R_{\mathrm{pore}})$.
Therefore, to a first approximation, controlling the surface charge can mimic controlling the electrical potential, so the results of this study can be informative regarding the case of field effect nanofluidic transistors too. 

Using an electrode to control the electrical potential near the nanopore leads to the presence of dielectric interfaces between materials of different polarization properties (electrolyte vs.\ metal, for example).
Polarization charges are induced at these dielectric boundaries that are different in every configuration of the ions, therefore, their presence influences the outcome of the calculations through influencing the probabilities of the individual configurations.
Calculation of induced charges or the electrical potential produced by them is a time consuming process compared to the homogeneous dielectric model and precalculated applied potential used here, because  the ion-ion interactions are not additive any more \cite{boda-pre-69-046702-2004,boda_2006,boda-prl-98-168102-2007}.
We refer studying this important case to future studies.

It is common to include electrodes (through imposing Dirichlet boundary conditions) and dielectric boundaries in mean field calculations (such as PNP).
These calculations, however, include the effect of polarization charges only on the \textit{average} electrical potential.
Electrostatic correlations resulting from the effect of induced charges on individual ionic configurations is ignored.
If the electrodes are far from the nanopore, this approximation can be sufficient, however. 

\subsection*{Comparison of PNP and NP+LEMC}
\label{subsec:sigpot}

One of the motivations of this work was to produce results for the model nanopore transistor using both a mean-field continuum theory (PNP) and a hybrid method including particle simulations (NP+LEMC) that can compute ion size effects and electrostatic correlations beyond the mean-field treatment.
In the light of the results we can conclude that PNP is able to capture the qualitative behavior of the device as shown by Figs.\ \ref{Fig2}--\ref{Fig8}. 

This indicates that the behavior of ionic profiles (as the first-order determinant of current) mainly depends on the interaction of ions with pore charges and applied field, while interaction of ions beyond interaction with the mean electric field is secondary.
Interaction with pore charges tunes the depth of depletion zones and directly modulates the electric current.
The applied potential makes the profiles asymmetric along the axial dimension and produces the driving force of the steady-state current.

The approximate nature of the PNP theory appears in quantitative disagreement between PNP and NP+LEMC results.
This can be seen both in the current data (Figs.\ \ref{Fig2}A, \ref{Fig3}, \ref{Fig4}A, \ref{Fig5}A, \ref{Fig6}A, \ref{Fig7}A, and \ref{Fig8}) and in the concentration profiles (Figs.\ \ref{Fig2}B, \ref{Fig5}B, \ref{Fig6}B, and \ref{Fig7}B). 
Sources of this quantitative disagreement are the following. 
(1) Ions modeled as point charges in PNP can approach the charged  wall infinitely close, so their interaction with the surface charge is overestimated near the wall.
PNP, therefore, overestimates concentration profiles at the peaks (Figs.\ \ref{Fig2}B and \ref{Fig5}B).
(2) Lack of hard sphere exclusion in PNP also tends to cause overestimation compared to NP+LEMC.
(3) Lack of electrostatic correlations in PNP, on the other hand, tends to decrease concentration profiles in the depletion zones compared to NP+LEMC, where ions that have peaks (counterions) tend to draw the ions of opposite charges (coions) into the depletion zones through pair-correlations.

Qualitative agreement, however, indicates that PNP is a proper tool to study the behavior of this system and those even larger in dimensions as demonstrated by several computational studies \cite{daiguji_nl_2005,kalman_am_2008,ai_sab_2011,singh_jap_2011b,jiang_pre_2011,singh_lc_2012,pardon_acis_2013,volkov_l_2014,yeh_sab_2015,Singh_PCCP_2017}.
Basically, PNP works well for these systems (also in the case of bipolar diodes), becase the behavior of these devices is primarily driven by the depletion zones caused by mean-field effects (interaction with pore charges and applied field).

Qualitative disagreement is expected in cases where ionic correlations casue asymmetic behavior  such as electrolytes containing multivalent ions (e.g., 2:1 and 3:1 electrolytes).
Furthermore, PNP cannot compute cases where the size of ions and specific interactions with binding sites are important such as in the case of sensors \cite{madai-jcp-147-244702-2017}. 
In general, particle simulations are better suited for modeling sensors based on specific interactions and geometries.

\subsection*{Comparison of PNP studies from literature}
\label{subsec:pnpliterature}

The behavior of the device studied here depends on all the parameters used in the model.
One basic reason of the fact that our model shows reasonable switching behavior and considerable current response to changing $\sigma_{\mathrm{x}}$ for such short pores is that we use relatively large surface charge.
Its value, $\pm 1 e/\mathrm{nm}^{2}$, however, is typical in experiments for PET nanopores.
The other reason is that we use small $R_{\mathrm{pore}}/\lambda_{\mathrm{D}}$ values.
If this ratio is small (narrow pore or low concentration), the double layer formed at the pore wall does not reach a bulk behavior in the pore center and depletion zones of coions can form in the respective regions.

A similar parameter domain was considered by Gracheva et al.\ \cite{gracheva_acsnano_2008,Gracheva_JCE_2008,nikolaev_jce_2014} who considered a nanopore through a semiconductor membrane and by Park et al.\ \cite{park_mfnf_2015} who considered a model called double-well nanofluidic channel.
The qualitative behavior reported by these authors is similar to the model considered by us.
Extended depletion zones created by gate voltages of appropriate sign were found.

Other studies considered wider and longer pores at lower concentrations keeping the $R_{\mathrm{pore}}/\lambda_{\mathrm{D}}$ ratio close to 1 \cite{daiguji_nl_2005,singh_jap_2011b,singh_lc_2012,Singh_PCCP_2017}.
Surface charge was quite low in these studies that did not make it possible to exclude the coions from the pore.
Depletion zones, therefore, were not formed in the entire positively or negatively charged regions of the nanopore.
Whether the scaling behavior found in our model holds in this parameter regime is unclear.
This question will be addressed in later studies.

Instead, a different mechanism worked that produced the depletion zones at the junctions of the differently charged regions.
This behavior can be observed in Fig.\ \ref{Fig4}B (solid red curve).
The concentration of cation is made asymmetric by the applied field.
The profile decreases in the central x region from $z\sim -4$ nm to $z\sim 4$ nm reaching quite a low value at $z\sim 4$ nm at the junction of the x and the right n regions.
This junctional depletion zone can be made deeper by making the pore longer or the voltage larger.
Fig.\ 4(c) of Daiguji et al.\ \cite{daiguji_nl_2005}, for example, clearly shows this behavior for 0.015 $e/\mathrm{nm}^{2}$ surface charge, 5 mM concentration, 2 $\mu$m x region, and 5 V voltage.

\section{Summary}
\label{sec:summary}

To summarize, we identified two different mechanisms for creating depletion zones depending on the values of nanopore parameters, especially, the magnitudes of the surface charges ($\sigma_{0}$).
If $\sigma_{0}$ is large (our case),  depletion zones are formed in the entire n and x regions (not only at the junctions) if the $R_{\mathrm{pore}}/\lambda_{\mathrm{D}}$ ratio is low enough (overlapping double layers and excluded coions).
In this case, interaction with the surface charges ($\sigma_{\mathrm{x}}$ and $\sigma_{\mathrm{n}}$) is the primary effect that creates the depletion zones in the radial dimension (see the insets of Fig.\ \ref{Fig5}B).
This mechanism works even if the pore is short and voltage is low.

If surface charge is low enough and/or  $R_{\mathrm{pore}}/\lambda_{\mathrm{D}}$ is large enough, bulk electrolytes are formed at the centerline of the pore so coions are not excluded.
Depletion zones for \textit{both} ions at one of the junctions are created by the applied field along the axial dimension.
The two mechanisms can combine \cite{cheng_acsnano_2009}.
The second mechanism, for example, can enhance the effect of the first mechanism as pore length increases as  seen in Fig.\ \ref{Fig4}.

We identified a scaling behavior of device function ($I_{\mathrm{ON}}/I_{\mathrm{OFF}}$) in relation to the $R_{\mathrm{pore}}/\lambda_{\mathrm{D}}$ ratio.
This scaling works for both PNP and NP+LEMC.
The two methods provide qualitatively similar results indicating that device behavior is governed by mean-field effects (interaction with mean potential produced by surface charges, applied field, and all the ions).

\section*{Acknowledgements}
\label{sec:ack}

We gratefully acknowledge  the financial support of the National Research, Development and Innovation Office -- NKFIH K124353. 
Present article was published in the frame of the project GINOP-2.3.2-15-2016-00053.
BM acknowledges financial support from the Austrian Academy of Sciences \"OAW via the New Frontiers  Grant NST-001. 
Supported by the \'{U}NKP-17-4 New National Excellence program of the Ministry of Human Capacities (MV).


\end{document}